 \documentclass[smallabstract,smallcaptions]{dccpaper}

\usepackage{epsfig}
\usepackage{citesort}
\usepackage{amsmath}
\usepackage{amssymb}
\usepackage{color}
\usepackage{url}
\usepackage{float}
\usepackage{subfig}
\usepackage{algorithm} 
\usepackage{algpseudocode} 
\usepackage{comment}
\usepackage[useregional]{datetime2}

\usepackage{xspace}
\usepackage{multicol}

\usepackage{todonotes}

\newlength{\figurewidth}
\newlength{\smallfigurewidth}
\usepackage{booktabs}

\usepackage{tikz} 
\usetikzlibrary{positioning}
\usetikzlibrary{shapes.arrows}
\usepackage{xcolor}
\usepackage[normalem]{ulem}

\tikzset{
    mybox/.style={rectangle,
        draw,
        rounded corners,
        minimum width=1cm,
        inner sep=5pt,
        align=center,
        minimum height=1cm
    },
    myarrow/.style={draw=black,
        fill=white,
        minimum width=0.2cm,
        single arrow
    },
    longarrow/.style={draw=none,
        shading=axis,
        left color=white,
        right color=blue,
        minimum width=0.6cm,
        single arrow,
        anchor=east
    }
}

\colorlet{darkred}{red!80!black}
\colorlet{darkblue}{blue!60!black}
\colorlet{darkgreen}{green!60!black}
\colorlet{darkgray}{white!50!black}

\newcommand{\dgreen}[1]{{\color{darkgreen} #1}}

\newcommand{\orange}[1]{{\color{orange} #1}}
\newcommand{\red}{\textcolor{red}}
\newcommand{\blue}{\textcolor{blue}}

\newcommand{\violet}{\textcolor{violet}}

\usepackage{caption}
\newcommand{\Suf}{\textrm{Suf}}

\newcommand{\BWT}{\ensuremath{\mathrm{BWT}}\xspace}

\newcommand{\SA}{\ensuremath{\mathrm{SA}}}

\newcommand{\lex}{\ensuremath{\mathrm{lex}}}
\newcommand{\col}{\ensuremath{\mathrm{colex}}}
\newcommand{\rev}{\ensuremath{\mathrm{rev}}}

\newcommand{\SAP}{\ensuremath{\mathrm{SAP}}}

\renewcommand\phi\varphi
\renewcommand\epsilon\varepsilon

\newcommand{\dol}{{\tt \$}}

\def\chr19{\texttt{chr19}}

\def\eBWT{\ensuremath{\mathrm{eBWT}}\xspace}

\newcommand*{\PlusPlus}{%
\kern0.3ex\raisebox{-0ex}{\sebox{0.8}{\kern-0.4ex+}}%
\kern-0ex\raisebox{0.5ex}{\sebox{0.8}{\kern-0.4ex+}}}

\begin{document}

\title
{\large
\textbf{Computing the optimal BWT of very large string collections}
}

\author{%
Davide Cenzato$^{\ast}$, Veronica Guerrini$^{\dag}$, Zsuzsanna Lipt\'ak$^{\ast}$, and Giovanna Rosone$^{\dag}$\\[0.5em]
{\small\begin{minipage}{\linewidth}\begin{center}
\begin{tabular}{ccc}
$^{\ast}$University of Verona, Italy & \hspace*{0.5in} & $^{\dag}$University of Pisa, Italy\\
\url{davide.cenzato@univr.it} && \url{veronica.guerrini@di.unipi.it}\\
\url{zsuzsanna.liptak@univr.it} && \url{giovanna.rosone@unipi.it}
\end{tabular}
\end{center}\end{minipage}}
}

\maketitle

\thispagestyle{empty}


\bigskip

\begin{abstract}
It is known that the exact form of the Burrows-Wheeler-Transform (BWT) of a string collection depends, in most implementations, on the input order of the strings in the collection.  Reordering strings of an input collection affects the number of equal-letter runs $r$, arguably the most important parameter of BWT-based data structures, such as the FM-index or the $r$-index.  Bentley, Gibney, and Thankachan [ESA 2020] introduced a linear-time algorithm for computing the permutation of the input collection which yields the minimum number of runs of the resulting BWT.  

In this paper, we present the first tool that guarantees a Burrows-Wheeler-Transform with minimum number of runs (optBWT), by combining {\it i)} an algorithm that builds the BWT from a string collection (either SAIS-based 
[Cenzato et al., SPIRE 2021] or BCR 
[Bauer et al., CPM 2011]); {\it ii)} the SAP array data structure introduced in [Cox et al., Bioinformatics, 2012]; and 
{\it iii)} the algorithm by  Bentley et al. 

We present results both on real-life and simulated data, showing that the improvement achieved in terms of $r$ with respect to the input order is significant and the overhead created by the computation of the optimal BWT negligible, making our tool competitive with other tools for BWT-computation in terms of running time and space usage. In particular, on real data the optBWT obtains up to 31 times fewer runs with only a $1.39\times$ slowdown.
\noindent
Source code is available at \url{ https://github.com/davidecenzato/optimalBWT.git}.
\end{abstract}

\Section{1. Introduction}

The {\em Burrows-Wheeler Transform} (BWT)~\cite{BW94} is a reversible text transformation that performs a symbol permutation of the input, resulting in a string which is often easier to compress than the original string.

Mantaci et al.\ in 2007~\cite{MRRS_TCS2007} defined the \eBWT\ ({\em extended BWT}), generalizing the \BWT\ from a single string to a string collection, by sorting the cyclic rotations of each input string according to the $\omega$-order, which differs from the usual lexicographic order. An important property of the \eBWT is that it is independent of the input order of the strings in the collection. However, it wasn't until 2021 that the first linear-time algorithms for constructing the \eBWT\ were presented~\cite{BannaiKKP21,BCLMR_SPIRE21_pfpebwt}, and only the latter has been implemented.
Possibly due to this, most tools for computing the~\BWT\ of string collections~\cite{BCR_TCS2013,HoltMcMilan2014,Li2014ropebwt,BVPPR_JCB2019,EMT_AMB2019_egap,BGKLMM_AMOB2019_BigBWT,LouzaTGPR20,DN_CPM2022_grlBWT} employ alternative definitions of the BWT which append an end-of-string symbol (often called {\em dollar}) to each input string.

The method presented in~\cite{BCR_CPM2011,BCR_TCS2013}  (with two different approaches, named BCR and BCRext) was the first to extend to a string collection ${\cal M}$ the \BWT defined by sorting the suffixes of the strings\footnote{Similarly to the case of a single string, when appending a different dollar to the strings in ${\cal M}$, the $\omega$-order coincides with the lexicographical order (a related study on the two order relations in~\cite{BonomoMRRS14}).} to which a different dollar is appended (making ${\cal M}$ an ordered set), without concatenating them. 
Otherwise, one could concatenate the input strings separating them with different dollars,
and sort the suffixes of the concatenation, with or without appending an additional end-of-string symbol $\#$.
See first three columns in Fig.~\ref{fig:runningEx}.
In order not to increase the size of the alphabet, usually the tools output the \BWT string using the same dollar, in spite of being implicitly distinct. 
In the systematic categorization of the different \BWT-variants according to the final output \BWT given in~\cite{CL_CPM2022}, this \BWT string is called {\em multidollar BWT} (fifth column in Fig.~\ref{fig:runningEx}) and it is shown to be dependent on the input order. In fact, if the same input collection is given with the strings in some permuted order, then the output can differ significantly.

Arguably the most important parameter of the BWT is its number of equal-letter runs,  commonly referred to as $r$.
Using runlength-encoding, the space requirement of \BWT-based data structures is proportional to $r$. 
There exist other types of BWT-variants that reduce the number of runs in the output, e.g., the authors in~\cite{GMRS_CPM2019} introduced a new family of BWT variants based on context adaptive alphabet orderings and on local orderings. However, the analysis of these BWT variants is beyond the scope of this paper, since we focus on the multidollar BWT, where the parameter $r$ is heavily affected by the input order. This was already remarked in~\cite{CoxBJR12}, where two different heuristics for reducing $r$ were introduced, the {\em rlo-heuristic} (called {\em colex} in~\cite{CL_CPM2022}), see also~\cite{Li2014ropebwt}, and the {\em sap-heuristic}.

The two  heuristics are obtained by permuting the symbols within special ranges, called SAP-intervals (\emph{same-as-previous}), intervals associated with suffixes equal up to the dollars. 
They can be represented along with the \BWT by means of the data structure {\em SAP-array}~\cite{CoxBJR12}.
Within SAP-intervals, one can permute the symbols by grouping them into as few runs as possible. 
It is easy to construct examples (see Fig.~\ref{fig:runningEx}) on which neither of these two heuristics results in a \BWT\ with minimal number of runs.

Bentley et al.~\cite{BentleyGT20} recently presented a linear-time algorithm that computes a permutation of the input collection minimizing $r$, but they gave no implementation. 

\SubSection{Contributions} 

In this paper, we give an implementation of the algorithm in~\cite{BentleyGT20} by means of the SAP-array~\cite{CoxBJR12}, 
for computing the \BWT with the minimum number of runs. We provide an on-the-fly construction of the SAP-array while building the \BWT of a string collection, using two different algorithms:
one is our adaptation\footnote{Preliminary version D.\ Cenzato and Zs.\ Lipt\'ak: {\em Computing the optimal BWT using SAIS.} presented at: 17th Workshop on Compression, Text, and Algorithms (WCTA 2022), Concepci\'on, Chile, 11 Nov.\ 2022.} of the SAIS-based algorithm of~\cite{BCLMR_SPIRE21_pfpebwt}, the other is the BCR-based  algorithm~\cite{BCR_CPM2011,BCR_TCS2013}.

Note that ours is the first tool that guarantees to output a \BWT of a string collection 
with minimal number of runs, in terms of reordering of input strings. 

This is significant not only because the storage space of most BWT-based data structures is proportional to $r$, such as RLBWT~\cite{RLBWT2009} or $r$-index~\cite{GagieNP18}, but also because it allows to use the minimum number of runs as a repetitiveness measure for string collections. 
As was pointed out in~\cite{CL_CPM2022}, the parameter $r$ should be standardized, since it is being increasingly used as a parameter of the dataset (string collection).  However, with the presence of different BWT variants which are all dependent on the input order, this parameter is not well-defined.

We performed several experiments both on simulated and real-life datasets. For each of these, we report the increase in the number of runs  of different input orderings with respect to the optimal BWT, showing that the improvement can be very significant. Moreover, our performance data show that the computational overhead is negligible, compared to computing the BWT given by the input order.

In particular, on real data the optBWT obtains up to 31 times fewer runs with only a $1.39\times$ slowdown, making our tool competitive with other tools for BWT-computation in terms of running time and space usage, while on simulated data we obtained a factor of up to 7.5 (with {\em P.\ aeruginosa}).
We are also interested in the behaviour of the number of runs of the optBWT in dependence of the read length. To this end, we fix a coverage and simulate Illumina reads of varying lengths. 


\begin{figure}[]
\tiny{
$$
\begin{array}{|c|@{\ }c|@{\ }c|c|c@{\ }|c@{\ }|c@{\ }|c@{\ }|c|l|}
\hline
\multicolumn{3}{|c|}{\mbox{multidollar BWT approaches}} &   & \multicolumn{4}{c}{\mbox{different ordering}} &   & \\
\hline
\mbox{$S_1\$_1 \cdot\cdot  S_5\$_5\#$} & \mbox{$\{S_1\$_1, ..,S_5\$_5\}$} & \mbox{$S_1\$_1 \cdot\cdot S_5\$_5$} & \mbox{SAP} &  \mbox{inputBWT} &  \mbox{dolEBWT} &  \mbox{colexBWT} & \mbox{sapBWT} &  \mbox{optBWT} & \mbox{sorted suffixes} \\
  \hline
\$_5 &      &      &            &    &     &           &                      &            & \#     \\
\hline
\text{A}&\text{A}&\text{A}& \textbf{0} &\text{A}&\text{T}&\text{A}&\text{A}& \dgreen{\textbf{T}} & \$     \\
\text{A}&\text{A}&\text{A}& \textbf{1} &\text{A}&\text{A}&\text{A}&\text{A}& \dgreen{\textbf{T}} & \$     \\
\text{T}&\text{T}&\text{T}& \textbf{1} &\text{T}&\text{T}&\text{T}&\text{T}& \dgreen{\textbf{T}} & \$     \\
\text{T}&\text{T}&\text{T}& \textbf{1} &\text{T}& \text{A}&\text{T}&\text{T}& \dgreen{\textbf{A}} & \$     \\
\text{T}&\text{T}&\text{T}& \textbf{1} &\text{T}&\text{T}&\text{T}&\text{T}& \dgreen{\textbf{A}} & \$     \\
\text{G}&\text{G}&\text{G}& \textbf{0} &\text{G}&\text{A}&\text{A}&\text{G}& \red{\textbf{A}} & \text{A}\$    \\
\text{A}&\text{A}&\text{A}& \textbf{1} &\text{A}&\text{G}&\text{G}&\text{A}& \red{\textbf{G}} & \text{A}\$    \\
\text{G}&\text{G}&\text{G}& 0 &\text{G}&\text{G}&\text{G}&\text{G}&\text{G}& \text{AA}\$   \\
\text{T}&\text{T}& \text{T} & \textbf{0} &\text{T}&\text{G}&\text{G}&\text{T}& \blue{\textbf{G}} & \text{CCT}\$  \\
\text{G}&\text{G}&\text{G}& \textbf{1} &\text{G}&\text{T}&\text{T}&\text{G}& \blue{\textbf{T}} & \text{CCT}\$  \\
\text{T}&\text{T}&\text{T}& 0  &\text{T}&\text{T}&\text{T}&\text{T}&\text{T}& \text{CGA}\$  \\
\text{C}&\text{C}& \text{C}&\textbf{0}&\text{C}&\text{C}&\text{C}&\text{C} & \orange{\textbf{T}} &\text{CT}\$   \\
\text{T}&\text{T}&\text{T}& \textbf{1} &\text{T}& \text{C} &\text{C}&\text{C} &\orange{\textbf{C}} & \text{CT}\$   \\
\text{C}&\text{C}& \text{C} & \textbf{1} & \text{C}  &\text{T}& \text{T} & \text{T}  &\orange{\textbf{C}} & \text{CT}\$   \\
\text{C}&\text{C}& \text{C} & 0 & \text{C} &\text{C} & \text{C} & \text{C} & \text{C} & \text{GA}\$   \\
\text{G}&\text{G}&\text{G}& 0 &\text{G}&\text{G}&\text{G}&\text{G}&\text{G}& \text{GAA}\$  \\
\$_4    &\$_5    &\$_4 & 0        & \$ & \$ & \$         & \$          & \$         & \text{GCCT}\$ \\
\$_1    & \$_2   &\$_1 & 0        & \$ & \$ & \$         & \$          & \$         & \text{GGAA}\$ \\
\text{C}&\text{C}&\text{C}&\textbf{0}& \text{C} & \text{C}& \text{C}   & \text{C} &\violet{\textbf{C}}  & \text{T}\$    \\
\text{C}&\text{C}&\text{C}& \textbf{1}&\text{C}& \text{C}& \text{C} & \text{C}  &\violet{\textbf{C}}& \text{T}\$    \\
\text{C}&\text{C}& \text{C}& \textbf{1} &\text{C}&\text{C}&\text{C}&\text{C}&\violet{\textbf{C}}&\text{T}\$    \\
\$_2&\$_3&\$_2&0 & \$ & \$ & \$  & \$ & \$ & \text{TCCT}\$ \\
\#   & \$_1 & \$_5 & 0          & \$ & \$ & \$ & \$    & \$  & \text{TCGA}\$ \\
\text{T}&\text{T}&\text{T}& 0  &\text{T}&\text{T}&\text{T}&\text{T}&\text{T}& \text{TCT}\$  \\
\$_3 & \$_4 & \$_3 & 0       & \$ & \$ & \$      & \$     & \$    & \text{TTCT}\$ \\
\hline
\multicolumn{4}{|r|}{\mbox{number of equal-letter runs}}       & 17 & 17 & 14         & 17     & 11         &    \\   
\hline
\end{array}
$$
}

\vspace{-4mm}

\caption{The output of different multidollar \BWT approaches (the resulting BWTs differ only in the dollars) applied to the collection 
${\cal M}=\{\tt TCGA,GGAA,TCCT,TTCT,GCCT\}$. The SAP-array (where the SAP-interval are in bold) and the BWT outputs with different re-ordering of ${\cal M}$.}
\label{fig:runningEx}

\end{figure}

\Section{2. Basics}

Let $\Sigma$ be a finite ordered alphabet of size $\sigma$. We use the notation $S=S[1..n]$ for a string of length $n$ over $\Sigma$, $S[i]$ for its $i$'th character, and $S[i..j]$ for the substring $S[i]\cdots S[j]$, for $i\leq j$. By convention $S[i..j] = \epsilon$ if $i>j$, where $\epsilon$ denotes the empty string. The length of string $S$ is denoted $|S|$.  Substrings of $S[1..n]$ of the form $S[1..i]$ are called {\em prefixes}, and substrings of the form  $S[i..n]$ {\em suffixes}; we denote the $i$'th suffix of $S$ by $\Suf_i(S)$. A substring (prefix, suffix) of $S$ is called {\em proper} if it does not equal $S$. A {\em rotation} (also called {\em conjugate}) of string $S$ is a string of the form $S[i..n]S[1..i-1]$, for some $1\leq i \leq n$. The {\em reverse} of string $S$ is denoted $S^{\rev} = S[n]\cdots S[1]$. Note that we index strings from $1$. 
The {\em Parikh vector} of a string $S$ is an integer vector
$(p_1,\ldots,p_{\sigma})$, where $p_j =|\{ i \mid S[i]={\tt a}_j\}|$ gives the multiplicity in $S$ of the $j$'th character ${\tt a}_j\in\Sigma$. A {\em run} in $S$ is a maximal substring consisting of the same character. 
For example, the string ${\tt CAAACCCTTTTG}$ has Parikh vector $(3,4,1,4)$ and $5$ runs. 

To mark the end of string $S$, often a new character  (usually denoted \dol) not belonging to $\Sigma$ is appended to it; $\dol$ is set to be smaller than all characters in $\Sigma$. When convenient, we simply write $S[n+1]=\dol$. 

The {\em lexicographic order} on strings is defined as $S<_{\lex} T$ if $S$ is a proper prefix of $T$, or if there exists an index $1\leq i$ such that $S[i]<T[i]$ and for all $j < i$, $S[j]=T[j]$. The {\em colexicographic order}, or {\em colex order}, is defined as $S<_{\col} T$ if $S^{\rev}<_{\lex} T^{\rev}$. (The colex order is sometimes referred to as {\em reverse lexicographic order}, or {\em rlo}, see~\cite{CoxBJR12,Li2014ropebwt}). 

Let $S$ be a string over $\Sigma$ and $S[n+1]=\dol$. The {\em suffix array}~\cite{MM93} \SA\ of $S$ is a permutation of the indices $1,2,\ldots, n+1$ such that $\SA[i]=j$ if $\Suf_j(S)$ is the $i$'th suffix of $S$ in lexicographic order among all suffixes of $S$. 
The {\em Burrows-Wheeler-Transform} (BWT)~\cite{BW94} of $S$ is defined as a permutation $L$ of the characters of $S$: $L[i] = \dol$ if $\SA[i] = 1$, and $L[i] = T[\SA[i]-1]$ otherwise.\footnote{Since we assume that $S$ is terminated by a \dol, this is equivalent to the alternative definition involving rotations given in~\cite{BW94}: $L[i]$ is the last character of the $i$'th rotation of $S$ in lexicographic order among all rotations.}

A {\em string collection} is a multiset of strings ${\cal M} = \{S_1,\ldots, S_k\}$, where each $S_i$ is assumed to be terminated by a different dollar-character $\dol_i$ and  $\dol_1 < \dol_2 < \ldots < \dol_k$.  Let $|S_i|=n_i$, then $||{\cal M}|| = \sum_i n_i + k$ is the total length of the collection. 

The \BWT\ of ${\cal M}$ can be defined as the classical \BWT\ of the concatenated string $S_1\$_1S_2\$_2\cdots S_k\$_k$.
Alternatively, it can be defined without concatenation as follows: Let $\Suf_t[S_j]$ be the $i$'th suffix in lexicographic order, among all suffixes of strings in ${\cal M}$, then $\BWT[i] = \$_j$ if $t=1$, and $\BWT[i]=S_j[t-1]$ otherwise. 


\Section{3. Algorithm for Computing the optBWT}

In this section, we describe the computation of the optBWT in two steps: $i)$ building an arbitrary BWT and its \SAP-array, $ii)$ determining the optBWT.

First we define the {\em SAP-array}~\cite{CoxBJR12}, a binary array of length $||{\cal M}||$: $\SAP[i] = 1$ if
and only if the symbol $\BWT[i]$ is associated with a suffix which is {\em same as} its {\em previous} suffix (up to the dollar) in the list of sorted suffixes.
An {\em \SAP-interval} $\BWT[b..e]$ is a maximal interval in \BWT such that $\SAP[i]=1$, for all $b<i\leq e$. 
%
%
%
\SAP-intervals which contain more than one character 
correspond to left-maximal shared suffixes, {which were called}
{\em interesting intervals} 
in~\cite{CL_CPM2022}. 
In this paper, we introduce the {\em reduced \SAP-array} obtained from the \SAP-array
by setting $\SAP_{red}[i]=0$, $b<i\leq e$, for any \SAP-interval $\BWT[b..e]$ which is a run of the same symbol (see Table~\ref{tab:tuples}).

We will first explain how to obtain optBWT from an arbitrary BWT and the \SAP-array (or equivalently, the reduced SAP-array). 
Then we describe how to obtain the \SAP-array during the BWT-construction using an adaptation of the SAIS-based BWT-algorithm of~\cite{BCLMR_SPIRE21_pfpebwt}, and finally, how to obtain the reduced \SAP-array during \BWT-construction with BCR~\cite{BCR_CPM2011,BCR_TCS2013}. In the following, we use the term {\em interesting intervals} to denote \SAP-intervals containing more than one character. 
Due to space restrictions, we only sketch the two algorithms for building the SAP-array here. 

\SubSection {3.1 Computing the optimal BWT using the \SAP-array} \label{subsectOpt}

\scalebox{0.75}{
\begin{minipage}{1.30\linewidth}
\begin{table}[H]
{
\begin{tabular}{|r@{\ }||c@{\ }c@{\ }c@{\ }c@{\ }c@{\ }c@{\ }c@{\ }c@{\ }c@{\ }c@{\ }c@{\ }c@{\ }c@{\ }c@{\ }c@{\ }c@{\ }c@{\ }c@{\ }|}
\hline
\text{inputBWT}   & {\tt AATATAA} & {\tt GAACT} & {\tt CT} & {\tt C} & {\tt \$} & {\tt GG} & {\tt C} & {\tt A} & {\tt \$} & {\tt \$} & {\tt \$} & {\tt T} & {\tt AC} & {\tt AA} & {\tt GG} & {\tt \$} & {\tt \$} & {\tt \$} \\ 
\cline{2-19}
tuples       & {\tt (A,T)} & {\tt (A,C,G,T)} & {\tt (C,T)} & {\tt (C)} & {\tt (\$)} & {\tt (G)} & {\tt (C)} & {\tt (A)} & {\tt (\$)} & {\tt (\$)} & {\tt (\$)} & {\tt (T)} & {\tt (A,C)} & {\tt (A)} & {\tt (G)} & {\tt (\$)} & {\tt (\$)} & {\tt (\$)} \\ 
\cline{2-19}
tuples opt & {\tt (T,A)} & {\tt (A,G,C,T)} & {\tt (T,C)} & {\tt (C)} & {\tt (\$)} & {\tt (G)} & {\tt (C)} & {\tt (A)} & {\tt (\$)} & {\tt (\$)} & {\tt (\$)} & {\tt (T)} & {\tt (C,A)} & {\tt (A)} & {\tt (G)} & {\tt (\$)} & {\tt (\$)} & {\tt (\$)} \\ 
\cline{2-19}
optBWT          & {\tt TTAAAAA} & {\tt AAGCT} & {\tt TC} & {\tt C} & {\tt \$} & {\tt GG} & {\tt C} & {\tt A} & {\tt \$} & {\tt \$} & {\tt \$} & {\tt T} & {\tt CA} & {\tt AA} & {\tt GG} & {\tt \$} & {\tt \$} & {\tt \$} \\ 
\cline{2-19}
SAP-array        & {\tt 0111111} & {\tt 01111} & {\tt 01} & {\tt 0} & {\tt 0}  & {\tt 01} & {\tt 0} & {\tt 0} & {\tt 0}  & {\tt 0}  & {\tt 0}  & {\tt 0} & {\tt 01} & {\tt 01} & {\tt 01} & {\tt 0}  & {\tt 0}  & {\tt 0}  \\ 
\cline{2-19}
reduced SAP-a. & {\tt 0111111} & {\tt 01111}     & {\tt 01}    & {\tt 0}   & {\tt 0}    & {\tt 00}  & {\tt 0}   & {\tt 0}   & {\tt 0}    & {\tt 0}    & {\tt 0}    & {\tt 0}   & {\tt 01}    & {\tt 00}  & {\tt 00}  & {\tt 0}    & {\tt 0}    & {\tt 0}  \\ \hline
\end{tabular}
}
\caption{
The input \BWT\ and optBWT\ on the string collection ${\cal M} = \{{\tt TGA, CACAA, AGAGT, TAA, CGAGT, CCA, TA}\}$ together with their \SAP-array and reduced \SAP-array.
\label{tab:tuples}} 
\end{table}
\end{minipage}
}

\vspace{4mm}

It is clear that all characters of the \BWT are fixed except those within interesting intervals, and therefore, the \BWT\ can be varied only within these. 
In fact, the two heuristics employed in~\cite{CoxBJR12,Li2014ropebwt} reduce the number of runs {\em within} interesting intervals by grouping together all characters of the same type.
The algorithm of Bentley et al.~\cite{BentleyGT20} further reduces the number of runs by grouping together runs of the same character at {\em borders} of interesting intervals, wherever possible. The authors show that this can be modeled as a problem they refer to as {\em tuple ordering problem}, which in turn can be turned into a shortest path problem in a DAG.  
Each SAP-interval is mapped to a tuple containing those characters which occur in the interval at least once, while a position $i$ outside any SAP-interval with $\BWT[i]=c$ is mapped to $(c)$. See Table~\ref{tab:tuples} for an example.

We compute the optBWT in a single left-to-right scan of the input \BWT\ and the \SAP-array. As explained above, for every pair of neighboring \SAP-intervals, the goal is to place identical character runs on either side of the border. If more than one character is shared between the two intervals, then this choice is not unique. Note that this implies that both intervals are interesting. Moreover, which character has to be chosen may also depend on the other neighbors of the two intervals. Therefore, an arbitrary number of consecutive interesting intervals may have to be kept track of before the decision which characters to place at the borders can be made.

We maintain a stack to keep track of the Parikh vectors of the tuples for which the \BWT\ has not yet been output. 
For each new tuple, if the stack is empty, either we can output the \BWT\ immediately (see Algorithm~\ref{algo:opt} lines 2-3), or check if there exists a match with the last character output in the BWT. If so, we remove the character from the Parikh vector 
and output its occurrences (lines 5-7). 
Finally we place it in the stack (line 8).
Otherwise if the stack is not empty, we check whether the characters can now be assigned (lines 11-16). 
This is the case if the top Parikh vector in the stack shares 1 or 0 characters with the current one: if it is 1, then that character must be taken, otherwise an arbitrary character can be chosen. We can now empty the stack and write the corresponding parts of the \BWT. Finally, if some characters of the current Parikh vector were not written in the BWT, we place the remaining Parikh vector in the stack.

In Table~\ref{tab:tuples}, the \BWT\ starts with three interesting intervals. The corresponding Parikh vectors are placed on the stack.  Arriving at $i=16$, i.e.\ at the fifth 0 in the \SAP-array, the stack contains $(0,5,0,0,2), (0,2,1,1,1), (0,0,1,0,1)$. The current Parikh vector is $(0,0,1,0,0)$ (corresponding to {\tt C}), and {\tt C} is the only character in the intersection with the top Parikh vector $(0,0,1,0,1)$. Therefore,  the BWT corresponding to the three interesting intervals can now be output and the stack emptied: {\tt TTAAAAA|AAGCT|TC|C}, where we marked borders between interesting intervals by $|$. 
Note that if the symbols in the second interesting interval were permuted as {\tt AACGT}, then we would also get the minimal number of runs.


\begin{figure}[t]
\centering
\scalebox{0.7}{
\begin{minipage}{1\linewidth}
\begin{algorithm}[H]
	\caption{\label{algo:opt}Procedure to process a Parikh vector $P$}
	\begin{center}
	\begin{algorithmic}[1]
	  	    \If {Stack \textbf{is} empty}
                 \If {there is exactly one $j$ such that $P[j] > 0$} \hfill{ \textit{// interval not interesting} }
                    \State write $P[j]$ copies of character $j$                 
                 \Else
                    \If{ $P[x] > 0$ where $x$ is the last character inserted in the BWT }
                        \State write $P[x]$ copies of the character $x$, $P[x] \leftarrow 0$
                    \EndIf
                    \State Stack $\leftarrow $ \textit{pushTop}($P$)  \hfill{ \textit{// push a new Parikh vector on the stack} }
                \EndIf
            \Else
                \State $T \leftarrow $ Stack.\textit{top}()   \hfill{ \textit{// first element of the stack} }
                \If{there are at least two $j$ s.t.\ $T[j]>0$ and $P[j]>0$} 
                    \State Stack $\leftarrow $ \textit{pushTop}($P$) 
                \Else 
                	\State write corresponding characters for each $T$ in Stack  \hfill{\textit{// see text for details}}
                \EndIf
            \EndIf
	\end{algorithmic}
	\end{center}
\end{algorithm}
\end{minipage}
}
\end{figure}

\SubSection {3.2 Computing BWT and SAP using SAIS}

We generate the SAP-array during the computation of the BWT, using our adaptation of the SAIS-based algorithm of~\cite{BCLMR_SPIRE21_pfpebwt}. 
This is done by computing it in each recursion step and propagating it while mapping back one 
recursion level up. The SAP-array within a step can be computed along with the SA while inducing the L- and S-type suffixes. This is achieved via an adaptation of the inducing step that allows to propagate the information that we are within a shared suffix:  
Let $S_i[t..n_i]$ be a shared suffix; if at least two positions are preceded by the same character $c$ then $cS_i[t..n_i]$ corresponds to another SAP-interval. Since all occurrences of the same suffix are listed together in the SA, we can compute all SA-values in the new SAP-interval sequentially during the inducing step. This is carried out keeping track of suffixes starting with the same character, and updating the SAP-array accordingly in case they are induced by the same shared suffix. 

\SubSection {3.3 Computing BWT and SAP using BCR}

BCR algorithm is based on the idea of right-to-left scanning, at the same time, all the $k$ strings and building the \BWT through $\ell+1$ iterations, where $\ell$ is the length of the longest string. 
At each iteration, BCR considers a ``slice'' of (at most) $k$ characters from the strings:
it starts by concatenating the symbols preceding all $\dol_i$, for all $i$, building a {\em partial \BWT} ($\BWT_{0}$). Then, at iteration $j$, for $j=1,\ldots,k$, the symbols circularly preceding the suffixes $S_i[n_i-j+1..n_i]$ (for all $1\leq i\leq k$) are inserted in the partial $\BWT_{j-1}$ by simulating the insertion of these suffixes in the list of suffixes of length $h$ (for all $h<j$) lexicographically sorted.

During the $j$'th step, we are able to compute and propagate from one iteration to the next  the SAP-interval information (see also~\cite{BCR_CPM2011,BCR_TCS2013,CoxBJR12}\footnote{Note that unlike~\cite{CoxBJR12}, we compute SAP-intervals for the current iteration.}). Indeed, when inserting symbols circularly preceding a shared suffix $S_i[n_i-j+1..n_i]$ (for some $i$),
we can deduce the length of the SAP-interval that these symbols form (i.e., their number). Furthermore, we can distinguish whether a SAP-interval is an interesting interval or not (i.e., the symbols form a equal-character run), so that we can incrementally build along with the \BWT both the SAP-array and the reduced SAP-array.

\Section{4. Experimental Result and Discussion}

In this section, we assess the performance of our tool, named {\tt optimalBWT}. 
It is arranged as a pipeline that runs the two steps described in the previous section and, for building the \BWT and the \SAP-array, provides two approaches: one is an adaptation of the SAIS-based algorithm of~\cite{BCLMR_SPIRE21_pfpebwt} that mainly works in internal memory, and the other is the BCR approach working in semi-external memory. The choice of the approach depends on the resources available.

To evaluate the performance of {\tt optimalBWT},
we have designed a series of tests on both simulated and real-life short-read datasets (see Table~\ref{tab:datasets}). Tests were performed on a DELL PowerEdge R630 machine, $24$-core machine with Intel(R) Xeon(R) CPU E5-2620 v3 at $2.40$ GHz, with $128$ GB of internal memory.
\newline
\scalebox{0.75}{
\begin{minipage}{1.28\linewidth}
\begin{table}[H]
\centering
\begin{tabular}{ |l|l|l|r|r|r|r|r|r| } 
 \hline
 \multicolumn{2}{|l|}{dataset}  & description & \textrm{length \BWT $n$}  & \textrm{len.} & 
 \text{no.\ seq} 
 & \textrm{$r{_{opt}}$} & \textrm{$n/r{_{opt}}$} 
 & \textrm{$n/r$} \\ 
 \hline\hline 
 1&ERR732065--70 & {\em HIV-virus} 
 & 1,345,713,812 & 150 & 8,912,012  & 11,539,661 & 116.62 & 27.62 \\ 
 \hline
 2&SRR12038540 & {\em SARS-CoV-2 RBD} 
 & 1,690,229,250 & 50& 33,141,750  & 14,864,523 & 113.71 & 8.08 \\ 
 \hline  
 3&ERR022075\_1 & {\em E. Coli str.\ K-12} & 2,294,730,100 & 100& 22,720,100 & 71,203,469 & 
 32.23 &8.83  \\ 
 \hline
 4&SRR059298 & {\em Deformed wing virus} & 2,455,299,082 & 72 & 33,634,234 & 48,376,632 & 50.75 & 9.83\\
 \hline 
 5&SRR065389--90 &{\em C.\ Elegans}& 14,095,870,474 & 100& 139,563,074  & 921,561,895 & 
 15.30 &6.26  \\  
 \hline
 6&SRR2990914\_1 & {\em Sindibis virus} 
 & 15,957,722,119 & 36 & 431,289,787 & 105,250,120 & 151.62 & 4.81  \\  
 \hline
 7&ERR1019034 & {\em H.\ Sapiens} 
 & 123,506,926,658 & 100 & 1,222,840,858 & 10,860,229,434 & 11.37 & 5.35  \\  
 \hline
\end{tabular}
\caption{\label{tab:datasets}
Real-life datasets used in the experiments
together with the number of runs ($r_{opt}$) and the average runlength ($n/r_{opt}$) of the optBWT compared to the average runlength ($n/r$) of the inputBWT.
}
\end{table}
\end{minipage}
}

\vspace{4mm}

We compare the number of runs in the optBWT with respect to the input order (inputBWT), the lexicographic order (dolEBWT) and the two heuristics, {\em rlo-heuristic} (colexBWT) and {\em sap-heuristic} defined in~\cite{CoxBJR12} (sapBWT) (see also Fig.~\ref{fig:runningEx}).
Both these heuristics reduce the number of runs within interesting intervals by grouping together all characters of the same type: the {\em rlo-heuristic} achieves this implicitly, since by sorting the input strings in colexicographic order, identical characters are grouped together within each interesting interval. 
The {\em sap-heuristic} can be thought of as an approximation of the rlo-heuristic, in which the permutation of symbols within interesting intervals occurs during the on-the-fly construction of the \BWT (through BEETL-BCRext) and the SAP-array information is implicitly obtained by computing a SAP status (more details in \cite{CoxBJR12}).

For the real-life datasets, we show in Table~\ref{tab:real_perc_increment} the {\em factor increase} and the {\em percentage increase}\footnote{obtained by $\frac{r-r_{opt}}{r_{opt}}\cdot100$, where $r$ is the number of the runs of  the BWT variant.} in number of runs with respect to the optBWT for several datasets of different size, composition and read length. 
We also report the time and memory peak to construct the optBWT from scratch by choosing the algorithmic approach which has the best trade-off performance between the two proposed.
We note that on the real datasets the increase of $r$ with respect to the optBWT is significant for all different read lengths and $n/r_{opt}$ values.
In particular, the two short-read datasets SRR2990914\_1 and SRR1203854, featuring high $n/r_{opt}$
, show $31.5$ and $14.07$ times fewer runs than the input order \BWT spending only a $1.39\times$ and $1.15\times$ overhead in time when using the BCR- and SAIS-based approaches, respectively. On the other hand, on the large human dataset~\cite{Mallick2016} (122.3 Gb) even if the factor is smaller than the others, the $r$ saved is still over 10 billion with only a $1.48\times$ time overhead.

\begin{figure*}[ht!]
    \centering
    \subfloat{
 	\centering
 		\includegraphics[width=0.49\textwidth]{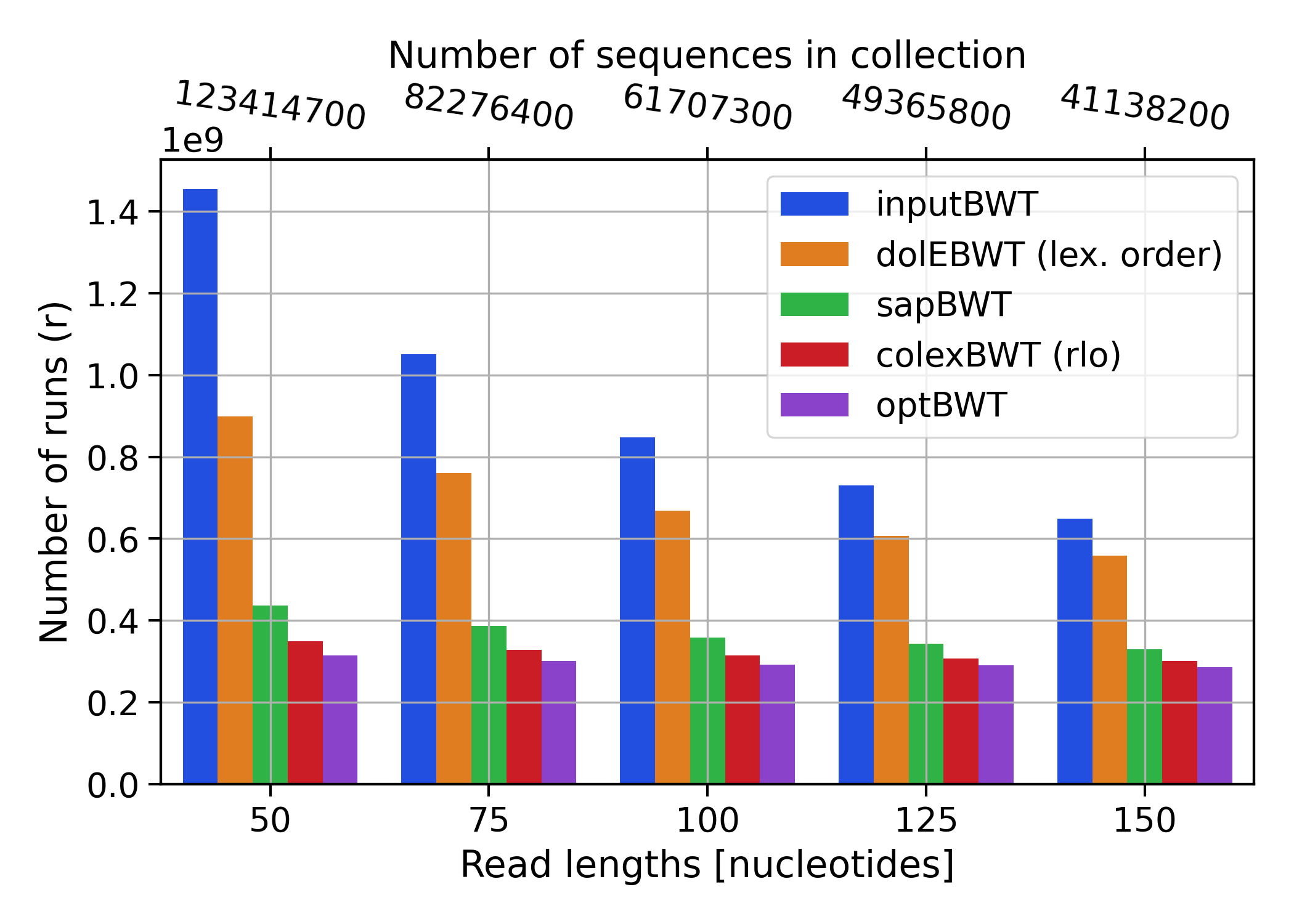}
    }~
   \subfloat{
     \centering
     		\includegraphics[width=0.49\textwidth]{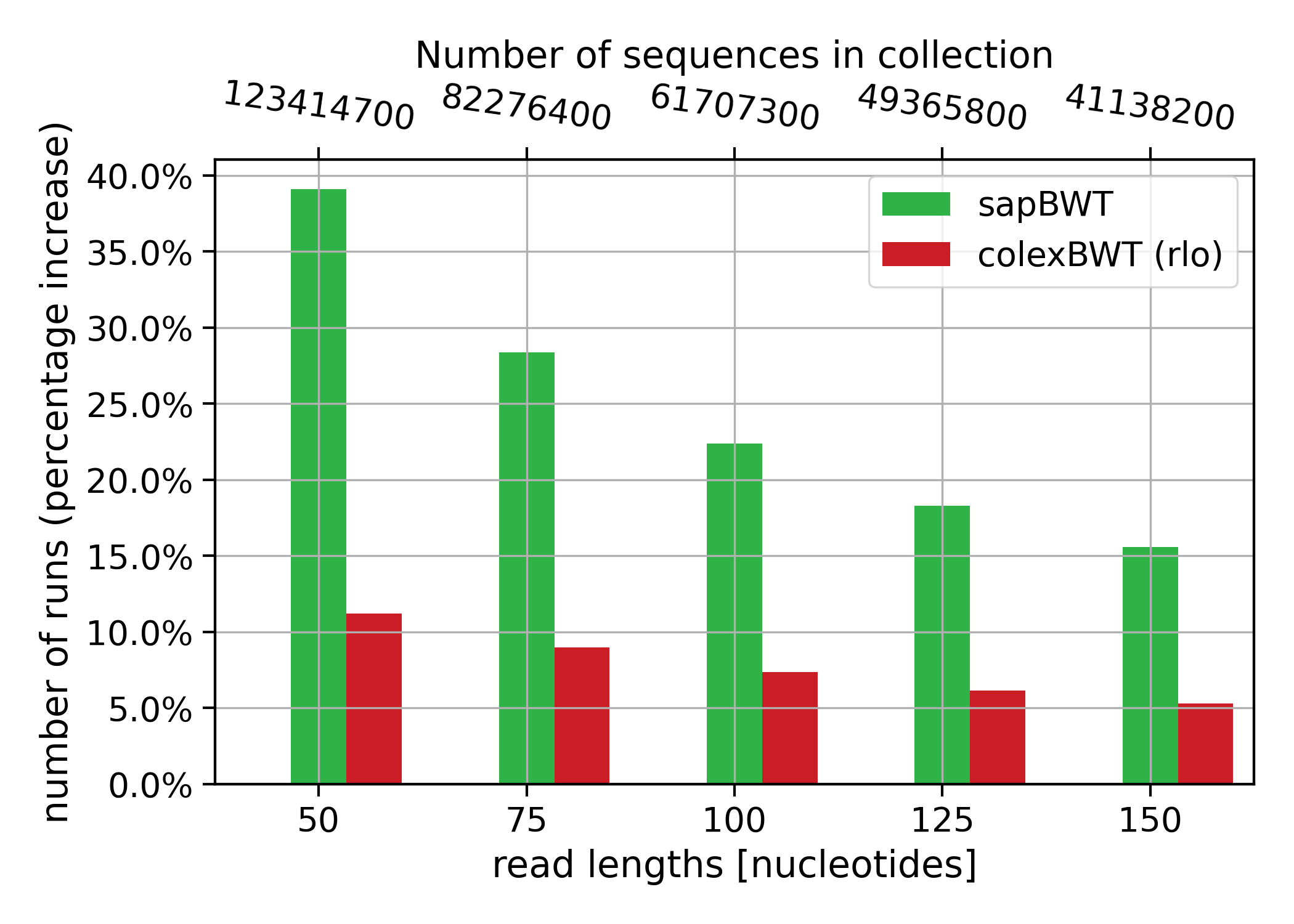}
    }
    \quad
    \subfloat{
 	\centering
 		\includegraphics[width=0.49\textwidth]{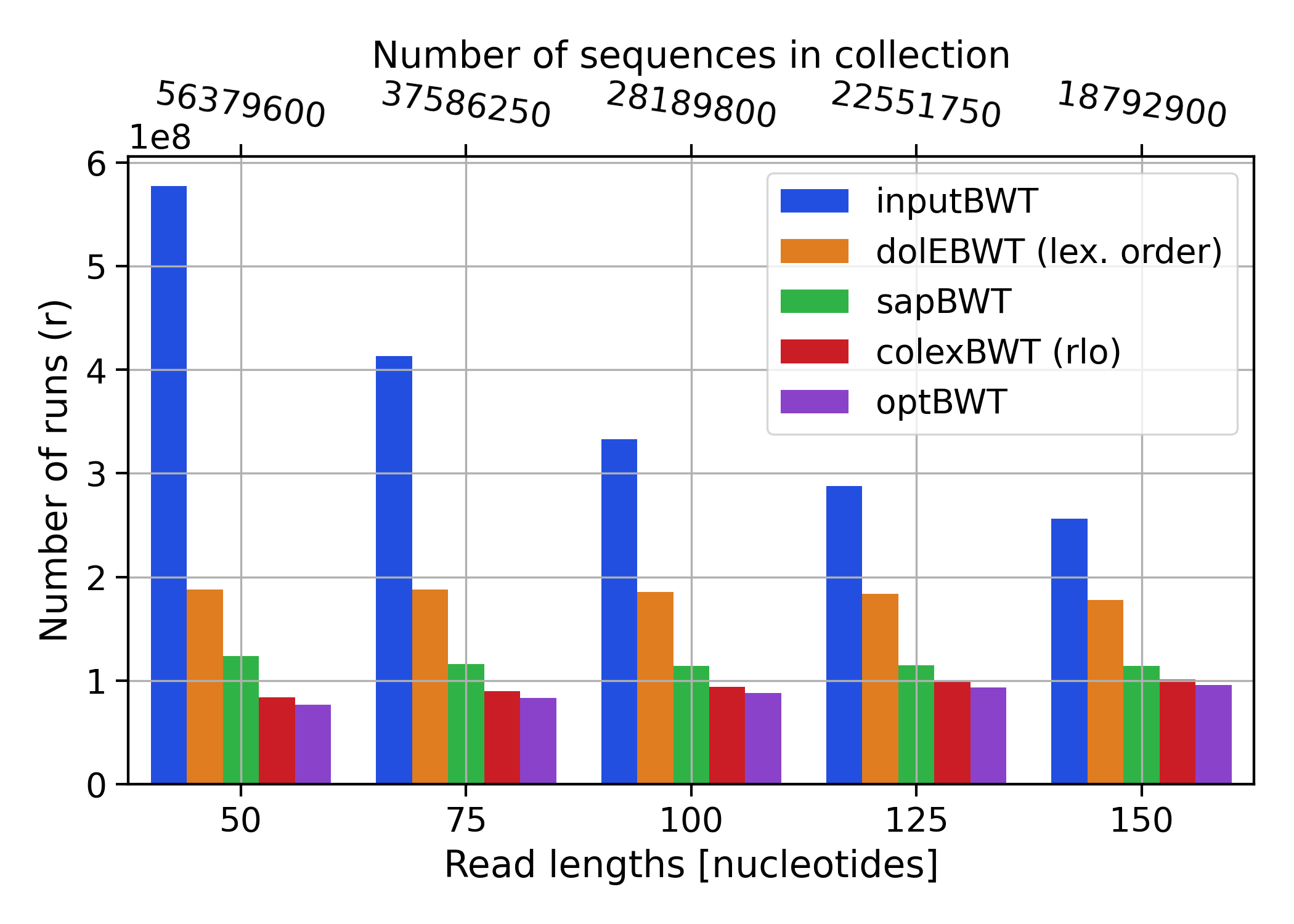}
    }~
   \subfloat{
     \centering
     		\includegraphics[width=0.49\textwidth]{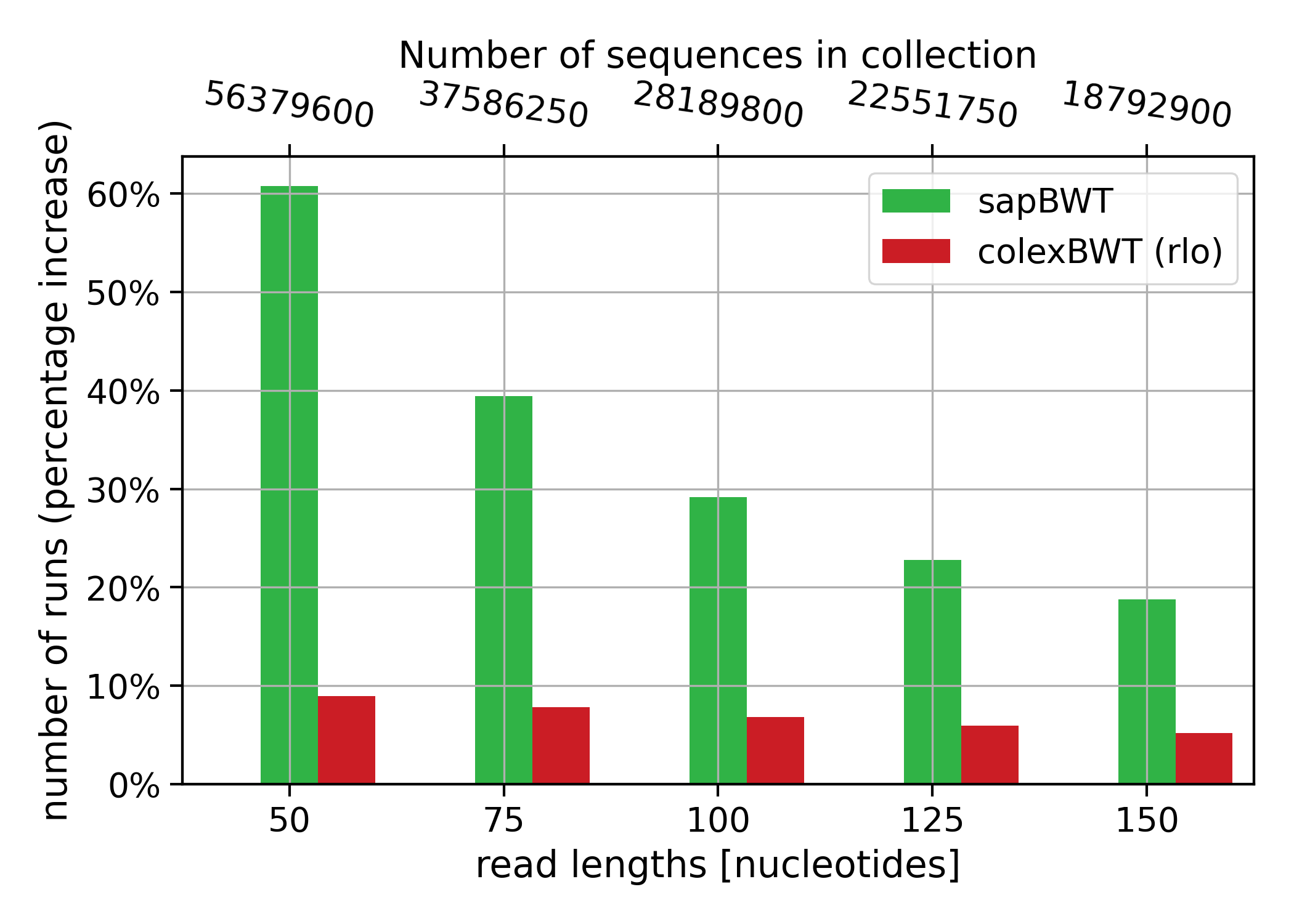}
    }
    \quad
    \subfloat{
 	\centering
 		\includegraphics[width=0.49\textwidth]{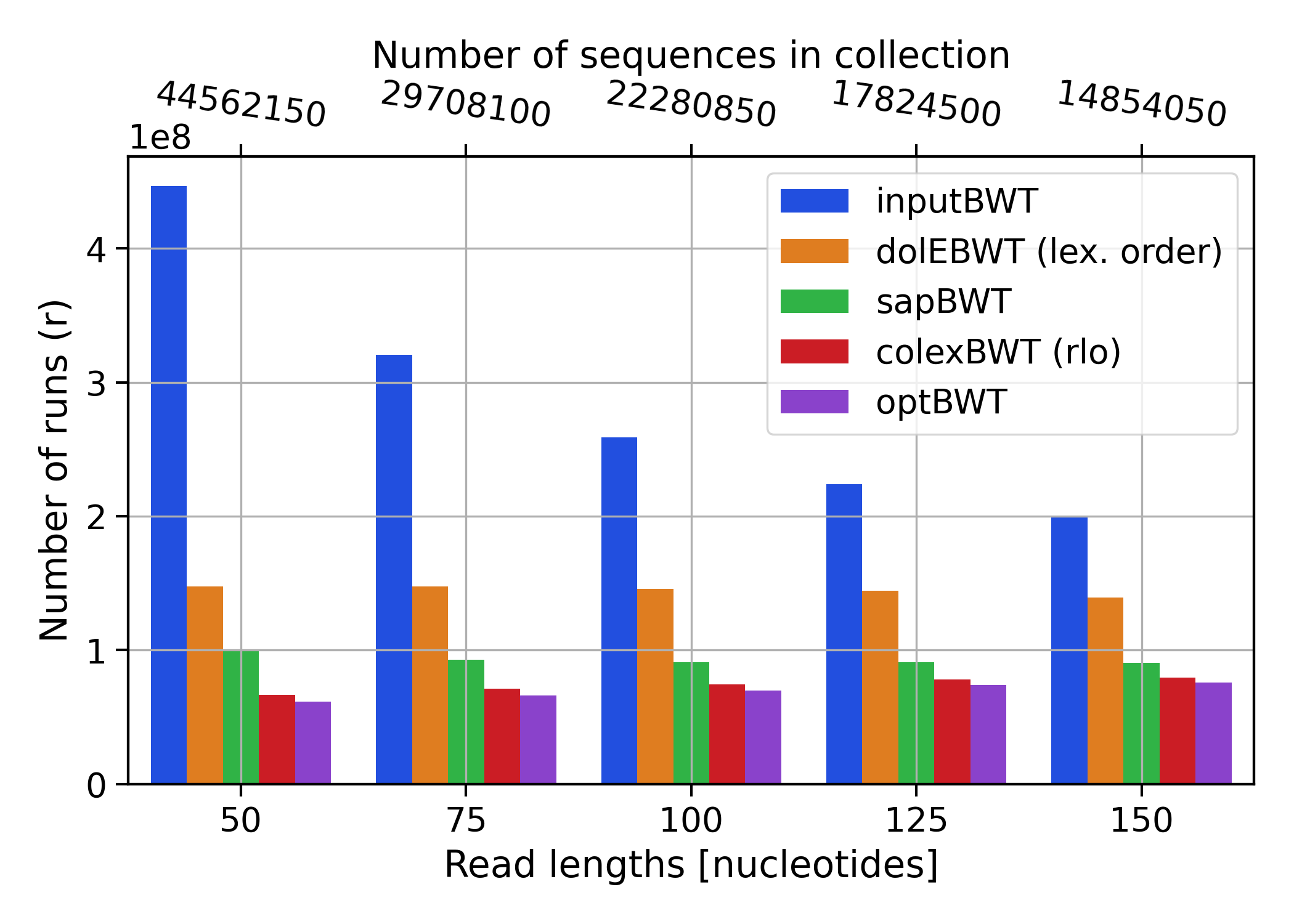}
    }~
   \subfloat{
     \centering
     		\includegraphics[width=0.49\textwidth]{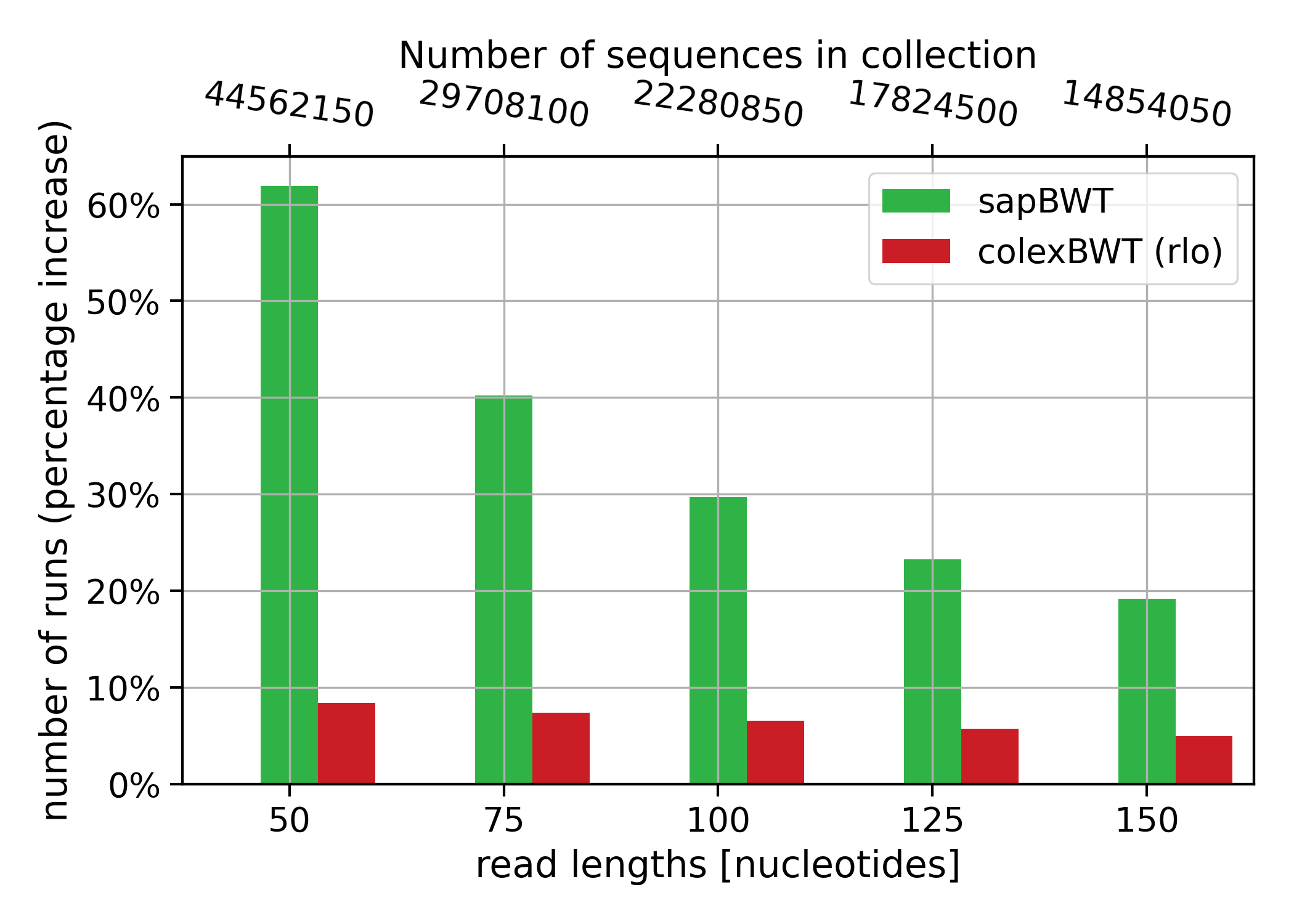}
    }
    \caption{Top to bottom: results regarding the number of runs on three simulated datasets of H. Sapiens chr.19 (cov.\ 100x), P. Aeruginosa (cov.\ 450x) and Salmonella (cov.\ 450x) varying read lengths. Left: number of runs. Right: percentage increase of sapBWT and colexBWT with respect to the optimal BWT.}

    \label{fig:bars}
\end{figure*}

For simulating short reads by varying read lengths, we used ART\footnote{\url{https://www.niehs.nih.gov/research/resources/software/biostatistics/art}}~(sequencing machine Illumina HiSeq 2500)
and various sequences (CP068259.2 \emph{H. Sapiens 
chr.19}, NC\_002516.2 \emph{P.\ Aeruginosa PAO1}, NC\_003197.2 \emph{S.\ enterica}).
In Fig.~\ref{fig:bars}, we plot the number of runs in the BWT variants while increasing the read length and keeping constant coverage (thus reducing the number of sequences).
As expected, by increasing the length of reads from 50 to 150 the number of runs decreases, as the number of reads and of the permutable symbols in the interesting intervals decreases.
However, the factor increase still is substantial for datasets with longer sequences, and the overhead to compute the optBWT is negligible for all read lengths (see also Table~\ref{tab:sim_time_mem}).\newline
\scalebox{0.75}{
\begin{minipage}{1.35\linewidth}
\begin{table}[H]
\centering
\begin{tabular}{ |l||r|r|r|r||r|r| } 
 \hline
 \text{data} & \multicolumn{4}{c||}{number of runs increase compared to optimal BWT} & \multicolumn{2}{c|}{resource usage}  \\ 
\cline{2-7}
\text{set}& \textrm{inputBWT}& \textrm{colexBWT (rlo)} & \textrm{sapBWT} & \textrm{dolEBWT} 
& RAM (GB) & Time (hh:mm:ss)\\
 \hline\hline
 1 & {\bf 4.22} \hspace{0.70mm} (322.26\%) & {1.03} \hspace{0.70mm} (3.48\%) & 1.53 (53.06\%) & {1.30} \hspace{0.70mm} (30.13\%) & {6.45 }($1.02\times$) & 7:18 ($1.12\times$)\\
 \hline
  2 & {\bf 14.07} (1306.95\%) & 1.15 (14.54\%) & 1.21 (20.75\%) & 3.52 (252.39\%) & 8.08 ($1.03\times$) & 6:32 ($1.15\times$)\\ 
 \hline
  3 & {\bf 3.65} \hspace{0.70mm} (264.90\%) & 1.07 \hspace{0.70mm} (6.52\%) & 1.30 (29.63\%) & 2.07 (107.01\%) & 11.15 ($1.04\times$) & 18:29 ($1.26\times$)\\ 
  \hline
 4 & {\bf 5.17 } \hspace{0.65mm}(416.52\%) & {1.04 } \hspace{0.65mm}(4.38\%) & 1.55 (55.33\%) & {1.55} \hspace{0.70mm} (54.87\%) & 21.03 ($1.02\times$) & 22:08 ($1.08\times$)
\\ \hline
  5 & {\bf 2.44} \hspace{0.70mm} (144.36\%) & 1.05 \hspace{0.70mm} (5.05\%) & 1.16 (15.73\%) & 2.03 (103.35\%) & 4.31 ($1.04\times$) & 2:25:46 ($1.28\times$)\\ 
  \hline
  6 & {\bf 31.49} (3048.66\%) & {1.04} \hspace{0.70mm} (4.30\%) &1.79 (79.40\%) &{1.89} \hspace{0.70mm} (89.17\%)& 8.86 ($1.05\times$)&1:59:46 ($1.39\times$)
  \\
  \hline
  7 & {\bf2.13} \hspace{0.5mm} (112.56\%) & 1.04 \hspace{0.5mm} (4.17\%) & 1.12 (11.89\%) & 1.96 \hspace{0.5mm} (96.04\%) &34.42 ($1.03\times$) & 26:24:18 ($1.48\times$)\\ 
  \hline
\end{tabular}
\caption{\label{tab:real_perc_increment} 
Results on the number of runs increase compared to the optBWT and resource usage. For each BWT variant we report the increase factor and the percentage increase  (in brackets). Total overhead in time and memory for building the optBWT from scratch with respect to the inputBWT is shown in brackets. For the first four datasets we used the SAIS-based approach, and the BCR-based one for the last three.
}
\end{table}
\end{minipage}
}
\scalebox{0.75}{
\begin{minipage}{1.35\linewidth}
\begin{table}[H]
\centering
\begin{tabular}{|l|r|r|r|r|r|r|r|}
\hline
             dataset & len. & {no. seq} & \multicolumn{1}{r|}{{no. runs}} & \multicolumn{2}{c|}{{RAM (GB)}} & \multicolumn{2}{c|}{{time (mm:ss)}}
             \\\cline{5-8}
             &&&{increase}&\multicolumn{1}{c|}{{BCR-based}} & \multicolumn{1}{c|}{{SAIS-based}} & \multicolumn{1}{c|}{{BCR-based}} & \multicolumn{1}{c|}{{SAIS-based}}\\
              \hline
NC\_002516.2 & 50 &    56,379,600
 & \textbf{7.50} (650.20\%) & 1.02 ($1.05\times$) & 13.91 ($1.03\times$) & 25:13 ($1.64\times$)  & 25:45 ($1.12\times$) \\\cline{2-8}
{\em P. aeruginosa} & 75 &   37,586,250 & \textbf{4.96} (395.76\%) & 0.98 ($1.03\times$)& 13.84 ($1.04\times$) & 27:39 ($1.58\times$) & 25:59 ($1.13\times$)\\\cline{2-8}
              & 100&    28,189,800  & \textbf{3.78} (277.91\%)& 0.52 ($1.05\times$)& 13.82 ($1.04\times$) & 30:39 ($1.54\times$)&  26:13 ($1.17\times$) \\\cline{2-8}
              & 125& 22,551,750 & \textbf{3.08} (208.18\%) & 0.51 ($1.04\times$) & 13.83 ($1.04\times$) & 34:14 ($1.57\times$) &  26:33 ($1.16\times$)\\\cline{2-8}
              & 150 & 18,792,900  & \textbf{2.67} (167.48\%) & 0.51 ($1.03\times$) & 13.83 ($1.04\times$) & 36:26 ($1.50\times$) & 26:32 ($1.18\times$)\\\cline{2-8}
              \hline
\end{tabular}
\caption{\label{tab:sim_time_mem} 
Results on the number of runs increase factor (percentage increase in brackets) compared to the optBWT, and resource usage for simulated datasets. Overhead in time and memory for building the optBWT from scratch using both approaches with respect to the inputBWT is shown in brackets.}
\end{table}
\end{minipage}
}
\Section{References}
\bibliographystyle{abbrv}
\bibliography{main.bib}

\end{document}